\documentclass[preprint,showpacs,preprintnumbers,amsmath,amssymb]{revtex4}

\usepackage{graphicx}
\usepackage{dcolumn}
\usepackage{bm}

\newcommand{\be}{\begin{equation}}
\newcommand{\en}{\end{equation}}
\newcommand{\bea}{\begin{eqnarray}}
\newcommand{\ena}{\end{eqnarray}}

\newcommand\rf[1]{(\ref{eq:#1})}
\newcommand\lab[1]{\label{eq:#1}}
\newcommand\nonu{\nonumber}
\newcommand\br{\begin{eqnarray}}
\newcommand\er{\end{eqnarray}}
\newcommand\ee{\end{equation}}

\newcommand\foot[1]{\footnotemark\footnotetext{#1}}
\newcommand\lb{\lbrack}
\newcommand\rb{\rbrack}

\renewcommand\({\left(}
\renewcommand\){\right)}

\newcommand\bc{\begin{center}}
\newcommand\ec{\end{center}}

\newcommand\partder[2]{\frac{{\partial {#1}}}{{\partial {#2}}}}

\renewcommand\a{\alpha}

\renewcommand\d{\delta}

\newcommand\eps{\epsilon}
\newcommand\vareps{\varepsilon}

\newcommand\G{\Gamma}

\newcommand\h{\frac{1}{2}}
\renewcommand\k{\kappa}
\renewcommand\l{\lambda}

\newcommand\m{\mu}
\newcommand\n{\nu}

\renewcommand\O{\Omega}

\newcommand\vp{\varphi}
\renewcommand\P{\Phi}
\newcommand\pa{\partial}

\renewcommand\th{\theta}

\newcommand{\ct}[1]{\cite{#1}}
\newcommand{\bib}[1]{\bibitem{#1}}

\newcommand\PRL[3]{\textsl{Phys. Rev. Lett.} \textbf{#1} (#2) #3}

\newcommand\PRD[3]{\textsl{Phys. Rev.} \textbf{D#1} (#2) #3}

\newcommand\PLB[3]{\textsl{Phys. Lett.} \textbf{#1B} (#2) #3}
\newcommand\CQG[3]{\textsl{Class. Quantum Grav.} \textbf{#1} (#2) #3}

\newcommand\IJMPD[3]{\textsl{Int. J. Mod. Phys.} \textbf{D#1} (#2) #3}

\newcommand\MPLA[3]{\textsl{Mod. Phys. Lett.} \textbf{A#1} (#2) #3}

\newcommand\vpdot{\stackrel{.}{\varphi}}

\newcommand\adot{\stackrel{.}{a}}
\newcommand\addot{\stackrel{..}{a}}

\begin{document}

\title{ Curvaton reheating {\bf mechanism} in a Scale Invariant Two Measures Theory }

\author{Eduardo I. Guendelman}
 \email{guendel@bgu.ac.il}
\affiliation{Physics Department, Ben Gurion University, Beer Sheva, Israel}
\author{Ram\'on Herrera}
\email{ramon.herrera@ucv.cl} \affiliation{Instituto de
F\'{\i}sica, Pontificia Universidad Cat\'{o}lica de
Valpara\'{\i}so, Avenida Brasil 2950, Valpara\'{\i}so, Chile.}

\date{\today}

\begin{abstract}
The curvaton reheating mechanism in a  Scale Invariant Two Measures Theory
defined in terms of two independent non-Riemannian volume forms
(alternative generally covariant integration measure densities) on
the space-time manifold which are metric independent is studied.
The model involves two scalar matter fields, a dilaton, that
transforms under scale transformations and it will be used also as
the inflaton of the model and another scalar, which does not
transform under scale transformations and which will play the role
of a curvaton field. Potentials of appropriate form so that the
pertinent action is invariant under global Weyl-scale symmetry are
introduced. Scale invariance is spontaneously broken upon
integration of the equations of motion. After performing
transition to the physical Einstein frame we obtain: (i) For given
value of the curvaton field an effective potential for the scalar
field with two flat regions for the dilaton which allows for a
unified description of both early universe inflation as well as of
present dark energy epoch;(iii) In the  phase corresponding to the
early universe, the curvaton has a constant mass and can oscillate
decoupled from the dilaton and that can be responsible for both
reheating and perturbations in the theory. In this framework, we obtain 
some interesting constraints on different parameters that appear in our model;
 (iii) For a definite
parameter range the model possesses a non-singular ``emergent
universe'' solution which describes an initial phase of evolution
that precedes the inflationary phase.
Finally we discuss
generalizations of the model, through the effect of higher
curvature terms,  where inflaton and curvaton can have coupled
oscillations.

\end{abstract}


\maketitle
\section{Introduction}
Inflationary universe models \cite{early-univ} have solved  some problems
of the Standard Hot Big Bang scenario, like the horizon and flatness problems.
High among the important accomplishments  of inflation are a natural source of the primordial perturbations \cite{primordial}. Inflation represents a
period of exponential accelerated expansion  and it so happens that the present Universe is also undergoing  a period of accelerated expansion \ct{accel-exp} ,\ct{accel-exp-2}, although much more slowly.
The possibility of continuously connecting an inflationary phase to a slowly
accelerating universe through the evolution of a single scalar field -- the
{\em quintessential inflation scenario} -- has been first studied in
Ref.\ct{peebles-vilenkin}. Also, 
$F(R)$ models can yield
both an early time inflationary epoch and a late time de Sitter phase with
vastly different values of effective vacuum energies \ct{starobinsky-2}.
For a recent proposal of a quintessential inflation mechanism based on
the k-essence \ct{k-essence} framework, see Ref.\ct{saitou-nojiri}. For
another recent approach to quintessential inflation based on the
``variable gravity'' model \ct{wetterich} and for extensive list of references
to earlier work on the topic, 
see Ref.\ct{murzakulov-etal}.

In a sequence of previous papers  \cite{TMTEMERGENTlast}, \cite{TMT-recent-1-a}, \cite{TMT-recent-1-c}
we approached the question of continuously connecting an inflationary phase to a slowly
accelerating universe through the evolution of a single scalar field in the context of the
Two Measures Theories \cite{TMT-orig-1}- \cite{susy-break}. Two Measures Theories models which use a non Riemannian measure
of integration in the action, and in the case of a scale invariant theory,  the scale invariance was spontaneously broken by
the equations of motion associated with the degrees of freedom which defined the non Riemanian  measure of integration in the action.
These degrees of freedom that define a non Riemannian measure of integration could be for example four scalar fields in four dimensions.
Models where four scalar fields in four dimensions have been used in the measure of integration and also in other parts of the action were
studied by Struckmeier \cite{Struckmeier}.  We also insisted in \cite{TMTEMERGENTlast}, \cite{TMT-recent-1-a}, \cite{TMT-recent-1-c}
(also see \cite{TMT-recent-1-b} for for a Two Measures Theory that
allows an emergent universe scenario, although without unification of inflation and dark energy)
in solving the initial singularity problem by having an Emergent scenario. Emergent scenario \cite{emergent-univ}
are non singular cosmological type of scenarios, where the universe starts as an Einstein Universe before developing
into the inflationary period. In this context, in the context of scale invariant two measures theories, in our most recent
 paper  \cite{TMTEMERGENTlast}, we used two non Riemannian measure of integration in the action, the equations of motion
 of each of the two measures of integration leading to two independent integration constants, these integration constants break
  scale invariance, and define the strength of the dark Energy density in the present universe, while they play no role in the early Universe  \cite{TMTEMERGENTlast}.

One problem with the scenarios that connect smoothly connect an inflationary phase to a slowly
accelerated phase, is that such models are not oscillating and therefore reheating may be problematic.
One solution to this is to introduce another field, the "curvaton" \cite{ureña}, \cite{ref1u}. The possible
 role of curvaton reheating in Non -Oscillatory Inflationary Models has been in particular
  studied by \cite{Feng-li,Campuzano:2005qw}. We will study the dynamic of the curvaton field
through different scenarios, and obtain the constraints upon
the free parameters on our model in order to have a feasible
curvaton stage during the reheating of the universe\cite{Feng-li,yo}. Firstly
we will consider that the curvaton coexists
with the inflaton field during the inflationary scenario, here
the inflaton energy density is the dominant component, and
the curvaton energy density should survive  to the  expansion of
the universe,  in which the curvaton field
 has to be effectively massless.
The next scenario the curvaton field gets
effectively massive during the kinetic epoch.  During this epoch, the curvaton
should oscillates in the minimum of its effective potential,  and
its energy density develops  as non-relativistic matter. Finally
the curvaton field decay into radiation,
and then the standard big bang cosmology is recuperated.
 At this point, we will study  two scenarios for the decay of the curvaton field,
since  the curvaton field could decays before or
after it becomes the dominant energy density of the universe.  However,
the curvaton field  introduces   an interesting  study  for  the observed
large-scale adiabatic density perturbations in the early universe. The hypothesis
of the curvaton field  suggests  that the adiabatic density perturbation proceeds  from the
curvaton  and
not from the inflaton field. In this framework, the adiabatic density perturbation is
originated
only after the inflationary scenario, and then the initial condition
 are purely isocurvature
perturbations. Recently the curvaton field 
is applied to the different theories\cite{Now}.
In this form, we will study the curvaton perturbation for both decays before or
after it becomes the dominant energy density of the universe.

In this paper we will see that the two measures theory that was discussed in \cite{TMTEMERGENTlast}
allows a simple generalization with the addition of a curvaton field as we will
 see in the next section.

 The outline of the paper goes as follow:
in Sec. II we give a  description of  two independent non-Riemannian volume-forms,
with the dilaton-inflaton and curvaton fields. In Section III the
curvaton field is analyzed  in the kinetic epoch.  The Section IV describes the curvaton decay
after its domination. The Section V explains  the decay of the curvaton field before it
dominates, and in
 Section VI includes
our conclusions.

\section{The Model: Two Independent Non-Riemannian Volume-Forms,
with the Dilaton-Inflaton and Curvaton fields}
We follow the general structure of the paper \cite{TMTEMERGENTlast}, but now we will enrich the field content
of the theory with a new field $\sigma$ which will not transform under scale transformations, so we write,
\label{TMMT}
\begin{equation}
S = \int d^4 x\,\P_1 (A) \Bigl\lb R + L^{(1)} \Bigr\rb +
\int d^4 x\,\P_2 (B) \Bigl\lb L^{(2)} + \eps R^2 +
\frac{\P (H)}{\sqrt{-g}}\Bigr\rb \; .
\lab{TMMT}
\end{equation}
Here the following notations are used:

\begin{itemize}
\item $\P_{1}(A)$ and $\P_2 (B)$ are two independent
non-Riemannian volume-forms, \textsl{i.e.}, generally covariant
integration measure densities on the underlying space-time
manifold: \be \P_1 (A) = \frac{1}{3!}\vareps^{\m\n\k\l} \pa_\m
A_{\n\k\l} \quad ,\quad \P_2 (B) = \frac{1}{3!}\vareps^{\m\n\k\l}
\pa_\m B_{\n\k\l} \; , \lab{Phi-1-2} \ee defined in terms of
field-strengths of two auxiliary 3-index antisymmetric tensor
gauge fields\foot{In $D$ space-time dimensions one can always
represent a maximal rank antisymmetric gauge field
$A_{\m_1\ldots\m_{D-1}}$ in terms of $D$ auxiliary scalar fields
$\phi^i$ ($i=1,\ldots,D$) in the form: $A_{\m_1\ldots\m_{D-1}} =
\frac{1}{D}\vareps_{i i_1\ldots i_{D-1}} \phi^i
\pa_{\m_1}\phi^{i_1}\ldots \pa_{\m_{D-1}}\phi^{i_{D-1}}$, so that
its (dual) field-strength $\P(A) = \frac{1}{D!}\vareps_{i_1\ldots
i_D} \vareps^{\m_1\ldots\m_D} \pa_{\m_1}\phi^{i_1}\ldots
\pa_{\m_D}\phi^{i_D}$.}. $\P_{1,2}$ take over the role of the
standard Riemannian integration measure density $\sqrt{-g} \equiv
\sqrt{-\det\Vert g_{\m\n}\Vert}$ in terms of the space-time metric
$g_{\m\n}$. \item $R = g^{\m\n} R_{\m\n}(\G)$ and $R_{\m\n}(\G)$
are the scalar curvature and the Ricci tensor in the first-order
(Palatini) formalism, where the affine connection $\G^\m_{\n\l}$
is \textsl{a priori} independent of the metric $g_{\m\n}$. Note
that in the second action term we have added a $R^2$ gravity term
(again in the Palatini form). Let us recall that $R+R^2$ gravity
within the second order formalism (which was also the first
inflationary model) was originally proposed in
Ref.\ct{starobinsky}. \item $L^{(1,2)}$ denote two different
Lagrangians of two scalar fields, the dilaton $\vp$, which will
play the role of an inflaton and now also the curvaton $\sigma$ .
The action will be taken of the form (similar to the choice in
Refs.\ct{TMT-orig-1}): \br L^{(1)} = -\h g^{\m\n} \pa_\m \vp
\pa_\n \vp -\h g^{\m\n} \pa_\m \sigma \pa_\n \sigma -\frac{\mu^2
\sigma^2}{2} \exp \{-\a\vp\} -  V(\vp)  , V(\vp) = f_1 \exp
\{-\a\vp\} \; ,
\lab{L-1} \\
L^{(2)} = -\frac{b}{2} e^{-\a\vp} g^{\m\n} \pa_\m \vp \pa_\n \vp + U(\vp)
 , U(\vp) = f_2 \exp \{-2\a\vp\} ,
\lab{L-2} \er where $\a, f_1, f_2, mu^2$ are dimensionfull positive
parameters, whereas $b$ is a dimensionless one. \item $\P (H)$
indicates the dual field strength of a third auxiliary 3-index
antisymmetric tensor gauge field: \be \P (H) =
\frac{1}{3!}\vareps^{\m\n\k\l} \pa_\m H_{\n\k\l} \; , \lab{Phi-H}
\ee whose presence is crucial for non-triviality of the model.
\end{itemize}

The scalar potentials have been chosen in such a way that the original action
\rf{TMMT} is invariant under global Weyl-scale transformations:
\br
g_{\m\n} \to \l g_{\m\n} \;\; ,\;\; \G^\m_{\n\l} \to \G^\m_{\n\l} \;\; ,\;\;
\vp \to \vp + \frac{1}{\a}\ln \l \;\;, \sigma  \to \sigma,
\nonu \\
A_{\m\n\k} \to \l A_{\m\n\k} \;\; ,\;\; B_{\m\n\k} \to \l^2 B_{\m\n\k}
\;\; ,\;\; H_{\m\n\k} \to H_{\m\n\k} \; 
\lab{scale-transf}
\er
For the same reason we have multiplied by an appropriate exponential factor
the scalar kinetic term in $L^{(2)}$ and also  $R$ and $R^2$ couple to the two
different modified measures because of the different scalings of the latter.

Let us note that the requirement about the global Weyl-scale symmetry
\rf{scale-transf} uniquely fixes the structure of the
non-Riemannian-measure gravity-matter action \rf{TMMT} (recall that the
gravity terms $R$ and $R^2$ are taken in the first order (Palatini) formalism).

Let us also note that the global Weyl-scale symmetry transformations defined in
\rf{scale-transf} are {\em not} the standard Weyl-scale (or conformal) symmetry
known in ordinary conformal field theory. It is straightforward to check that
the dimensionful parameters $\a, f_1, f_2$ present in \rf{L-1}-\rf{L-2}
do {\em not} spoil at all the symmetry given in \rf{scale-transf}.
In particular, unlike the standard form of the Weyl-scale transformation for
the metric the transformation of the scalar field $\vp$ is not the
canonical scale transformation known in standard conformal field theories.
In fact, as shown in the second Ref.\ct{TMT-orig-1} in the context of a
simpler than \rf{TMMT} model with only one non-Riemannian measure, upon
appropriate $\vp$-dependent conformal rescaling of the metric together with a scalar field
redefinition $\vp \to \phi \sim e^{-\vp}$, one can transform the latter
model into Zee's induced gravity model \ct{zee-induced-grav}, where its
pertinent scalar field $\phi$ transforms multiplicatively under the above
scale transformations as in standard conformal field theory.

The equations of motion resulting from the action \rf{TMMT} are as follows.
Variation of \rf{TMMT} w.r.t. affine connection $\G^\m_{\n\l}$:
\be
\int d^4\,x\,\sqrt{-g} g^{\m\n} \Bigl(\frac{\P_1}{\sqrt{-g}} +
2\eps\,\frac{\P_2}{\sqrt{-g}}\, R\Bigr) \(\nabla_\k \d\G^\k_{\m\n}
- \nabla_\m \d\G^\k_{\k\n}\) = 0,
\lab{var-G}
\ee
shows, following the analogous derivation in the Ref.\ct{TMT-orig-1}, that
$\G^\m_{\n\l}$ becomes a Levi-Civita connection:
\be
\G^\m_{\n\l} = \G^\m_{\n\l}({\bar g}) =
\h {\bar g}^{\m\k}\(\pa_\n {\bar g}_{\l\k} + \pa_\l {\bar g}_{\n\k}
- \pa_\k {\bar g}_{\n\l}\) \; ,
\lab{G-eq}
\ee
w.r.t. to the Weyl-rescaled metric ${\bar g}_{\m\n}$:
\be
{\bar g}_{\m\n} = (\chi_1 + 2\eps \chi_2 R) g_{\m\n} \;\; ,\;\;
\chi_1 \equiv \frac{\P_1 (A)}{\sqrt{-g}} \;\; ,\;\;
\chi_2 \equiv \frac{\P_2 (B)}{\sqrt{-g}} \; .
\lab{bar-g}
\ee

Variation of the action \rf{TMMT} w.r.t. auxiliary tensor gauge fields
$A_{\m\n\l}$, $B_{\m\n\l}$ and $H_{\m\n\l}$ yields the equations:
\be
\pa_\m \Bigl\lb R + L^{(1)} \Bigr\rb = 0 \quad, \quad
\pa_\m \Bigl\lb L^{(2)} + \eps R^2 + \frac{\P (H)}{\sqrt{-g}}\Bigr\rb = 0
\quad, \quad \pa_\m \Bigl(\frac{\P_2 (B)}{\sqrt{-g}}\Bigr) = 0 \; ,
\lab{A-B-H-eqs}
\ee
whose solutions read:
\be
\frac{\P_2 (B)}{\sqrt{-g}} \equiv \chi_2 = {\rm const} \;\; ,\;\;
R + L^{(1)} = - M_1 = {\rm const} \;\; ,\;\;
L^{(2)} + \eps R^2 + \frac{\P (H)}{\sqrt{-g}} = - M_2  = {\rm const} \; .
\lab{integr-const}
\ee
Here $M_1$ and $M_2$ are arbitrary dimensionfull integration constants and $\chi_2$
is an arbitrary dimensionless integration constant.

The first integration constant $\chi_2$ in \rf{integr-const} preserves
global Weyl-scale invariance \rf{scale-transf}, whereas
the appearance of the second and third integration constants $M_1,\, M_2$
signifies {\em dynamical spontaneous breakdown} of global Weyl-scale invariance
under \rf{scale-transf} due to the scale non-invariant solutions
(second and third ones) in \rf{integr-const}.

To this end let us recall that classical solutions of the whole set of equations
of motion (not only those of the scalar field(s)) correspond in the semiclassical
limit to ground-state expectation values of the corresponding fields.
In the present case some of the pertinent classical solutions
(second and third Eqs.\rf{integr-const}) contain arbitrary integration constants
$M_1,\, M_2$ whose appearance makes these solutions non-covariant w.r.t. the
symmetry transformations \rf{scale-transf}. Thus, spontaneous symmetry
breaking of \rf{scale-transf} is not necessarily originating from some fixed
extrema of the scalar potentials. In fact, as we will see in the next
Section below, the (static) classical solutions for the scalar field defined
through extremizing the effective Einstein-frame scalar potential
(Eq.\rf{U-eff} below) belong to the two infinitely large flat regions of the
latter (infinitely large ``valleys'' of ``ground states''), therefore, this does
not constitute a breakdown of the shift symmetry of the scalar field
\rf{scale-transf}. Thus, it is the appearance of the arbitrary integration
constants $M_1,\, M_2$, which triggers the spontaneous breaking of
global Weyl-scale symmetry \rf{scale-transf}.

Varying \rf{TMMT} w.r.t. $g_{\m\n}$ and using relations \rf{integr-const}
we have:
\be
\chi_1 \Bigl\lb R_{\m\n} + \h\( g_{\m\n}L^{(1)} - T^{(1)}_{\m\n}\)\Bigr\rb -
\h \chi_2 \Bigl\lb T^{(2)}_{\m\n} + g_{\m\n} \(\eps R^2 + M_2\)
- 2 R\,R_{\m\n}\Bigr\rb = 0 \; ,
\lab{pre-einstein-eqs}
\ee
where $\chi_1$ and $\chi_2$ are defined in \rf{bar-g},
and $T^{(1,2)}_{\m\n}$ are the energy-momentum tensors of the scalar
field Lagrangians with the standard definitions:
\be
T^{(1,2)}_{\m\n} = g_{\m\n} L^{(1,2)} - 2 \partder{}{g^{\m\n}} L^{(1,2)} \; .
\lab{EM-tensor}
\ee

Taking the trace of Eqs.\rf{pre-einstein-eqs} and using again second relation
\rf{integr-const} we solve for the scale factor $\chi_1$:
\be
\chi_1 = 2 \chi_2 \frac{T^{(2)}/4 + M_2}{L^{(1)} - T^{(1)}/2 - M_1} \; ,
\lab{chi-1}
\ee
where $T^{(1,2)} = g^{\m\n} T^{(1,2)}_{\m\n}$.

Using second relation \rf{integr-const} Eqs.\rf{pre-einstein-eqs} can be put
in the Einstein-like form:
\br
R_{\m\n} - \h g_{\m\n}R = \h g_{\m\n}\(L^{(1)} + M_1\)
+ \frac{1}{2\O}\(T^{(1)}_{\m\n} - g_{\m\n}L^{(1)}\)
\nonu \\
+ \frac{\chi_2}{2\chi_1 \O} \Bigl\lb T^{(2)}_{\m\n} +
g_{\m\n} \(M_2 + \eps(L^{(1)} + M_1)^2\)\Bigr\rb \; ,
\lab{einstein-like-eqs}
\er
where:
\be
\O = 1 - \frac{\chi_2}{\chi_1}\,2\eps\(L^{(1)} + M_1\) \; .
\lab{Omega-eq}
\ee
Let us note that \rf{bar-g}, upon taking into account second relation
\rf{integr-const} and \rf{Omega-eq}, can be written as:
\be
{\bar g}_{\m\n} = \chi_1\O\,g_{\m\n} \; .
\lab{bar-g-2}
\ee

Now, we can bring Eqs.\rf{einstein-like-eqs} into the standard form of Einstein
equations for the rescaled  metric ${\bar g}_{\m\n}$ \rf{bar-g-2},
\textsl{i.e.}, the Einstein-frame gravity equations:
\be
R_{\m\n}({\bar g}) - \h {\bar g}_{\m\n} R({\bar g}) = \h T^{\rm eff}_{\m\n},
\lab{eff-einstein-eqs}
\ee
with energy-momentum tensor corresponding (according to \rf{EM-tensor}):
\be
T^{\rm eff}_{\m\n} = g_{\m\n} L_{\rm eff} - 2 \partder{}{g^{\m\n}} L_{\rm eff},
\lab{EM-tensor-eff}
\ee
to the following effective Einstein-frame scalar field Lagrangian:
\be
L_{\rm eff} = \frac{1}{\chi_1\O}\Bigl\{ L^{(1)} + M_1 +
\frac{\chi_2}{\chi_1\O}\Bigl\lb L^{(2)} + M_2 + 
\eps (L^{(1)} + M_1)^2\Bigr\rb\Bigr\} \; .
\lab{L-eff}
\ee

In order to explicitly write $L_{\rm eff}$ in terms of the Einstein-frame
metric ${\bar g}_{\m\n}$ \rf{bar-g-2} we use the short-hand notation for the
scalar kinetic terms:
\be
X \equiv - \h {\bar g}^{\m\n}\pa_\m \vp \pa_\n \vp, Y \equiv - \h {\bar g}^{\m\n}\pa_\m 
\sigma \pa_\n \sigma,
\lab{X-def}
\ee
and represent $L^{(1,2)}$ in the form:
\be
L^{(1)} = \chi_1\O\, X + \chi_1\O\, Y -\frac{\mu^2 \sigma^2}{2} \exp \{-\a\vp\} - V \quad ,\quad L^{(2)} = \chi_1\O\,b e^{-\a\vp}X + U \; ,
\lab{L-1-2-Omega}
\ee
with $V$ and $U$ as in \rf{L-1}-\rf{L-2}.

From Eqs.\rf{chi-1} and \rf{Omega-eq}, taking into account \rf{L-1-2-Omega},
we find:
\be
\frac{1}{\chi_1\O} = \frac{(V + \frac{\mu^2 \sigma^2}{2} \exp \{-\a\vp\} -M_1)}{2\chi_2\Bigl\lb U+M_2 + \eps (V + \frac{\mu^2 \sigma^2}{2} \exp \{-\a\vp\}-M_1)^2\Bigr\rb}
\,\Bigl\lb 1 - \chi_2 \Bigl(\frac{b e^{-\a\vp}}{V + \frac{\mu^2 \sigma^2}{2} \exp \{-\a\vp\} -M_1} - 2\eps\Bigr) X\Bigr\rb \; .
\lab{chi-Omega}
\ee
At this point we see that keeping the $\epsilon$ contributions will lead to mixed curvaton - dilaton kinetic terms,
 i,e, $XY$ terms, these will lead to oscillations between these two fields. For simplicity, we want to stick to a more standard curvaton scenario,
 where the curvaton oscillates, while the inflaton (in our case the dilaton) does not participate in those oscillations. So we consider in what
  follows $\epsilon = 0$, in a future study the more complex case where $\epsilon \neq 0$ could 
  be studied .  Upon substituting expression \rf{chi-Omega} into \rf{L-eff} we arrive at the
explicit form for the Einstein-frame scalar Lagrangian:
\be
L_{\rm eff} = A(\vp, \sigma) X + B(\vp) X^2 + Y - U_{\rm eff}(\vp,  \sigma) \ ; ,
\lab{L-eff-final}
\ee
where:
\br
A(\vp, \sigma) \equiv 1 + \Bigl\lb \h b e^{-\a\vp}\Bigr\rb
\frac{V + \frac{\mu^2 \sigma^2}{2} \exp \{-\a\vp\} - M_1}{U + M_2 }
\nonu\\
= 1 + \Bigl\lb \h b e^{-\a\vp}  \Bigr\rb
\,\frac{f_1 e^{-\a\vp}+ \frac{\mu^2 \sigma^2}{2} \exp \{-\a\vp\} - M_1}{f_2 e^{-2\a\vp} + M_2 }
\; ,
\lab{A-def}
\er
and
\br
B(\vp) \equiv \chi_2 \frac{- \frac{1}{4} b^2 e^{-2\a\vp}}{U + M_2 }
\nonu\\
= \chi_2 \frac{ -\frac{1}{4} b^2 e^{-2\a\vp}
}{f_2 e^{-2\a\vp} + M_2 } \; ,
\lab{B-def}
\er
whereas the effective scalar field potential reads:
\be
U_{\rm eff} (\vp, \sigma) \equiv
\frac{(V+ \frac{\mu^2 \sigma^2}{2} \exp \{-\a\vp\} - M_1)^2}{4\chi_2 \Bigl\lb U + M_2\Bigr\rb}
= \frac{\(f_1 e^{-\a\vp}+ \frac{\mu^2 \sigma^2}{2} \exp \{-\a\vp\} -M_1\)^2}{4\chi_2\,\Bigl\lb
f_2 e^{-2\a\vp} + M_2 \Bigr\rb} \; ,
\lab{U-eff}
\ee
where in the last step the explicit form of $V$ and $U$ \rf{L-1}-\rf{L-2} are inserted.

Let us recall that the dimensionless integration constant $\chi_2$ is the
ratio of the original second non-Riemannian integration measure to the
standard Riemannian one \rf{bar-g}.

To conclude this Section let us note that choosing  the ``wrong'' sign of the scalar
potential $U(\vp)$ (Eq.\rf{L-2}) in the initial non-Riemannian-measure
gravity-matter action \rf{TMMT} is necessary to end up with the right sign in the
effective scalar potential \rf{U-eff} in the physical Einstein-frame effective
gravity-matter action \rf{L-eff-final}. On the other hand, the overall sign of the
other initial scalar potential $V(\vp)$ (Eq.\rf{L-2}) is in fact irrelevant since
changing its sign does not affect the positivity of effective scalar potential
\rf{U-eff}.

Let us also remark that the effective matter Lagrangian \rf{L-eff-final} is called
``Einstein-frame scalar Lagrangian'' in the sense that it produces the
effective energy-momentum tensor \rf{EM-tensor-eff} entering the effective
Einstein-frame form of the gravity equations of motion \rf{eff-einstein-eqs}
in terms of the conformally rescaled metric ${\bar g}_{\m\n}$ \rf{bar-g-2}
which have the canonical form of Einstein's gravitational equations. On the
other hand, the pertinent Einstein-frame effective scalar Lagrangian
\rf{L-eff-final} arises in a non-canonical ``k-essence'' \ct{k-essence} type
form.

A general remark concerning the counting of degrees of freedom is also in order. For 
this purpose it is crucial that generically the "first order formalism" where the connection
is a true independent degree of freedom and the  "second order formalism" where the connection
is assumed a priori to be the standard Christoffel  symbol are really generically different theories, 
except for the case of the Einstein-Hilbert action, so the
notion that these are just to different formulations of the same theory is justified only the case we
deal with the Einstein-Hilbert action, for generalizations, theories with a similar looking lagrangian 
are inequivalent in the two formulations, even the counting of degrees of freedom is different. For
example introducing non linear curvature terms or modified measures does not change the number of degrees of freedom in the first order formulation, where the new measure can be solved in terms of the Riemannian
times a function of the matter fields, this is not true if we were to consider the second order formalism.
In the same way, there is no increase in degrees of freedom by the introduction of non linear curvature terms in the first order formulation, like in our case the introduction of $R^2$ terms, again , this is not true in the second order formulation of the theory, where the introduction of non linear curvature terms does indeed changes the order of the equations and therefore causes an increase in degrees of freedom of the theory. In particular in our case, using the first order formulation,  the choice $\epsilon = 0$ does not represent a change in the number of degrees of freedom of the theory, just merely a particular parameter choice that simplifies the equations.

\section{Inflationary phase and Dark Energy phase from Flat Regions of the Effective Scalar Potential}
\label{flat-regions}

Depending on the sign of the integration constant $M_1$ we obtain two types of
shapes for the effective scalar potential $U_{\rm eff} (\vp)$ \rf{U-eff} . This sign determines
whether the effective potential has a zero or not.
The crucial feature of $U_{\rm eff} (\vp)$ is the presence of two infinitely large
flat regions, or more precisely  $\vp$ independent regions -- for large negative
and large positive values of the scalar field $\vp$.
For large negative values of $\vp$, which will characterize the inflationary region, we have for the effective potential and the
coefficient functions in the Einstein-frame scalar Lagrangian
\rf{L-eff-final}-\rf{U-eff}, when keeping  up to quadratic terms in $\sigma$ only :
\br
U_{\rm eff}(\vp) \simeq
\frac{f_1^2/f_2}{4\chi_2 } + \frac{m^2 \sigma^2}{2} \equiv \frac{f_1^2/f_2}{4\chi_2 } + U(\sigma); ,
\lab{U-minus} \\
A(\vp, \sigma) \simeq A_{(-)} \equiv 1+ \h b f_1/f_2 + \frac{ m^2 b f_1 \sigma^2}{4f_2} \;\; ,\;\;
B(\vp) \simeq B_{(-)} \equiv
- \chi_2 b^2/4f_2  \; .
\lab{A-B-minus}
\er
In the second flat region for large positive $\vp$, which will characterize the present slowly accelerated phase of the universe, we obtain:
\br
U_{\rm eff}(\vp) \simeq U_{(+)} \equiv
\frac{M_1^2/M_2}{4\chi_2 } \; ,
\lab{U-plus} \\
A(\vp) \simeq A_{(+)} \equiv 1 \quad ,\quad
B(\vp) \simeq B_{(+)} \equiv 0 \; ,
\lab{A-B-plus}
\er
where curvaton mass that appears in \rf{U-minus} in the first flat region in the minus region which is obtained is
\begin{equation}
m^2=\mu^2\frac{f_1}{2\chi_2 f_2}.
\label{m}
\end{equation}

This is the mass for the curvaton relevant to the reheating of the  universe.
Concerning the magnitude of the Dark Energy Density, if we take the integration
constant $\chi_2 \sim 1$, and
if we choose the scales of the scale symmetry breaking integration constants
$|M_1| \sim M^4_{EW}$ and $M_2 \sim M^4_{Pl}$, where $M_{EW},\, M_{Pl}$ are
the electroweak and Plank scales, respectively, we are then naturally led to
a very small vacuum energy density $U_{(+)}\sim M_1^2/M_2$ of the order:
\be
U_{(+)}\sim M^8_{EW}/M^4_{Pl} \sim 10^{-120} M^4_{Pl} \; ,
\lab{U-plus-magnitude}
\ee
which is the right order of magnitude for the present epoche's vacuum energy density.
In the present paper we will be mostly concerned with the $\varphi \rightarrow -\infty$ region.

It is interesting to think whether there is any strong motivation to choose $M_1 \sim M_{EW}^4$ and $M_2 \sim M_{Pl}^4$ for explaining the smallness of the Dark Energy Density. This kind of effect is generally
present in modified measure theories where the vacuum energy appears as some constant square divided by another constant as discussed first in the first two papers in ref. \citep{TMT-orig-1}.
What is suggestive about this type of equation is that it resembles the famous "see saw mechanism" used to obtain a small mass for the neutrino, not by fine tuning something to be very small, but rather, suppressing the neutrino mass by a big scale that enters as a denominator that enters in the expression of the diagonalized mass eigenvalues \cite{see saw}. So as in neutrino physics, the see-saw mechanism is widely employed to understand the tiny masses of the known neutrinos. In our case, we want to relate a see saw effect to understand the smallness of the cosmological constant. In the particle  physics case, small masses in the see saw mechanism, the crucial issue is the diagonalization of the mass matrix and in the case of our case the relevant process analogous  to the diagonalization of a mass matrix is the transition  to the Einstein frame,  where a see saw formula for the vacuum energy is obtained. Finally a natural choice for the choices for $M_1$ and $M_2$  must be determined in terms of the fundamental
mass scales  which we know are present in nature, which lead us naturally to the choice of 
 $M_1 \sim M_{EW}^4$ and $M_2 \sim M_{Pl}^4$ for explaining the smallness of the Dark Energy Density.

In this $\varphi \rightarrow -\infty$ region we will study the cosmological evolution.
To this end let us recall the standard Friedman-Lemaitre-Robertson-Walker
space-time metric :
\be
ds^2 = - dt^2 + a^2(t) \Bigl\lb \frac{dr^2}{1-K r^2}
+ r^2 (d\th^2 + \sin^2\th d\phi^2)\Bigr\rb,
\lab{FLRW}
\ee
and the associated Friedman equations
(recall the presently used units $G_{\rm Newton} = 1/16\pi$):
\be
\frac{\addot}{a}= - \frac{1}{12} (\rho + 3p) \quad ,\quad
H^2 + \frac{K}{a^2} = \frac{1}{6}\rho \quad ,\;\; H\equiv \frac{\adot}{a} \; ,
\lab{friedman-eqs}
\ee
describing the universe' evolution. Here:
\br
\rho = \h A(\vp, \sigma) \vpdot^2 + \h \dot{\sigma}^2 + \frac{3}{4} B(\vp) \vpdot^4 + U_{\rm eff}(\vp, \sigma) \; ,
\lab{rho-def} \\
p = \h A(\vp, \sigma) \vpdot^2 + \h \dot{\sigma}^2 + \frac{1}{4} B(\vp) \vpdot^4 -
 U_{\rm eff}(\vp, \sigma),
\lab{p-def}
\er
are the energy density and pressure produced by  the scalar fields $\vp = \vp (t)$ and $\sigma = \sigma(t)$.

The effective action for this cosmological "mini superspace" is
\br
S =\int dt a^3 p,
\er
\lab{p-action}
with $p$ given by \rf{p-def}
In the limit $\varphi \rightarrow -\infty$, $A$, $B$ and $ U_{\rm eff}$ become $\varphi$ independent
although some $\sigma$ dependence remains, therefore the above action with $p$ given by \rf{p-def} acquires
the symmetry  $\varphi \rightarrow \varphi + constant$, which means that the canonically conjugate momentum associated to $\varphi$ is a conserved quantity, that is
\begin{equation}
a^3(A\dot{\phi}+B\dot{\phi}^3)= C_k ,
\end{equation}
\label{constant}
where the quantity $C_k$ is a constant.
\section{Kinetic epoch}

During the  kinetic regime,  the dynamics of the Friedman-Robertson-Walker cosmology for
 our model becomes, taking the asymptotically  constant values of the coefficients $A$ and $B$,
given by \rf{A-B-minus}, neglecting also the $\sigma$ dependence of  $A$, then we have,
\begin{equation}
\ddot{\phi}[A+3B\dot{\phi}^2]+3H\dot{\phi}[A+B\dot{\phi}^2]=0,
\label{key_1}
\end{equation}
and
\begin{equation}
 6\,H^2\,=\rho_{\phi_k}\label{key_2},
\end{equation}
where the kinetic  energy density of the scalar  field, $\rho_{\phi_k}$,
is defined as

\begin{equation}
\rho_{\phi_{kin}}=\frac{A}{2}\dot{\phi}^2+\frac{3\,B}{4}\dot{\phi}^4.
\label{rhoi}
\end{equation}

We obtain a first integral of  Eq.~(\ref{key_1}) given by
\begin{equation}
A\dot{\phi}+B\dot{\phi}^3=\frac{C_k}{a^3}\,,\label{firin}
\end{equation}
where $C_k$ is an integration constant and is defined as
$$
C_k=a_k^3\,[A\dot{\phi_k}+B\dot{\phi_k}^3].
$$
We note that Eq.(\ref{firin}) agrees with the Noether conserved quantity defined before.
Here, the quantities
$\dot{\phi_k}$ and $a_k$
correspond to the  values at the beginning of the kinetic epoch for the
quantity $\dot{\phi}$ and the scale factor $a$,
respectively.

The real solution of Eq.(\ref{firin}) can be written as
\begin{equation}
\dot{\phi}=\dot{\phi}(a)=\frac{[9\tilde{B}a^{-3}+
\sqrt{12\tilde{A}^3+81\tilde{B}^2a^{-6}}\,\,]^{2/3}-2^{2/3}3^{1/3}\,\tilde{A}}{2^{1/3}3^{2/3}
\,[9\tilde{B}a^{-3}+
\sqrt{12\tilde{A}^3+81\tilde{B}^2a^{-6}}\;]^{1/3}},\label{sf}
\end{equation}
where
$\tilde{A}=\frac{A}{B}$, and $\tilde{B}=\frac{C_k}{B}$.

In this form, we can rewritten Eq.(\ref{sf}) as
\begin{equation}
\dot{\phi}=\dot{\phi_k}\,{\cal{F}}_a=\dot{\phi_k}\,{\cal{F}}(a/a_k),\label{mc}
\end{equation}
where the function ${\cal F}_a$ is given by
$$
{\cal F}_a={\cal F}(a/a_k)=\left(\frac{[9\tilde{B}a^{-3}+
\sqrt{12\tilde{A}^3+81\tilde{B}^2a^{-6}}\,\,]^{2/3}-2^{2/3}3^{1/3}\,\tilde{A}}
{2^{1/3}3^{2/3}\,[9\tilde{B}a^{-3}+
\sqrt{12\tilde{A}^3+81\tilde{B}^2a^{-6}}\;]^{1/3}}\right)\,\times
$$
$$
\left(\frac {2^{1/3}3^{2/3}\,[9\tilde{B}a_k^{-3}+
\sqrt{12\tilde{A}^3+81\tilde{B}^2a_k^{-6}}\;]^{1/3}}{[9\tilde{B}a_k^{-3}+
\sqrt{12\tilde{A}^3+81\tilde{B}^2a_k^{-6}}\,\,]^{2/3}-2^{2/3}3^{1/3}\,\tilde{A}}\right),
$$
such that ${\cal F}_a={\cal F}(a/a_k)\mid _{a=a_k}= 1$.

Now combining   Eqs.(\ref{rhoi}) and (\ref{mc}), we find an explicit relation for the
kinetic energy density in terms of the scale factor $a$
\begin{equation}
  \rho_{\phi_{kin}}=\frac{A}{2}\dot{\phi_k}^2{\cal F}_a^2+\frac{3B}{4}\dot{\phi_k}^4
  {\cal F}_a^4,\label{Impor}
\end{equation}
and
 the
Hubble parameter during the kinetic epoch can be written as
\begin{eqnarray}
 H_{kin}=\sqrt{\frac{\rho_{kin}}{6}}=\sqrt{\frac{1}{6}}\;\dot{\phi_k}{\cal F}_a\left[
 \frac{A}{2}+\frac{3B}{4}\dot{\phi_k}^2{\cal F}_a^2\right]^{1/2}.
 \label{H}
\end{eqnarray}

In the following we will analyze  the dynamic of the curvaton field, $\sigma$, through
different scenario. From these scenarios we will find some constraints of the
parameters in our model.

Initially , we consider that the energy density $\rho$, of the inflaton field, is the
dominant component when it is contrasted  with the curvaton energy density,
$\rho_\sigma$.
In the next scenario, the curvaton field $\sigma$ oscillates around the minimum of its effective potential
$U(\sigma)$.
 Its energy density developed  as a nonrelativistic matter and in the kinetic scenario,
the universe stays inflaton-dominated. In the last stage occurs   the decay of
the curvaton field into radiation, and therefore the  big-bang model is obtained.

For   the curvaton field $\sigma $ we
assumed  that obeys the Klein-Gordon
equation, in which
the  effective potential associates to  curvaton field
can be written as
$
U(\sigma)=\frac{m^2\sigma^2}{2}$,
where $m$ corresponds to the curvaton mass, see Eq.(\ref{m}).

During the inflationary epoch is assumed  that the curvaton mass
$m$ satisfied the condition $m\ll\,H_e$, where $H_e$ corresponds
to the Hubble factor at the end of inflation.  Recalled that the
inflationary evolution in our model is is described in detail in
Ref.\cite{TMTEMERGENTlast}.  In the inflationary scenario, the
curvaton field $\sigma$ would roll down its potential
 until its kinetic energy  vanished. In this situation the curvaton field
 assumes  a constant value, in which  $\sigma_*\approx\sigma_e$. In the following,
 the  subscript $*$
is refers to the epoch when the cosmological scales exit the horizon.

Following Ref.\cite{ureña}, we considered  that  during the kinetic epoch  the Hubble
factor decreases so that its value is similar to  the
curvaton mass, then $m\simeq H_{kin}$. In this way, considering  Eq.(\ref{H}), we
can written

\begin{equation}
m\simeq \sqrt{\frac{1}{6}}\;\dot{\phi_k}{\cal F}_{a_{m}}\left[
 \frac{A}{2}+\frac{3B}{4}\dot{\phi_k}^2{\cal F}_{a_{m}}^2\right]^{1/2},
\label{mh}
\end{equation}
or equivalently
$$\mu^2\frac{f_1}{\chi_2 f_2} \simeq \frac{1}{3}\;\dot{\phi_k}^2
{\cal F}_{a_{m}}^2\left[
 \frac{A}{2}+\frac{3B}{4}\dot{\phi_k}^2{\cal F}_{a_{m}}^2\right],
$$
where the `m' label corresponds to the quantities at the time during the kinetic epoch when
the curvaton mass is of the order of $H_{kin}$, and ${\cal F}_a\mid_{a=a_m}={\cal F}_{a_{m}}$ .

For avoiding  a period of curvaton-driven inflation, then we considered that
$\rho_{\phi_{kin}}|_{a_m}=\rho_{\phi_{kin}}^{(m)}\gg\rho_{\sigma}(\sim\,U(\sigma_e)
\simeq\,U(\sigma_*))$. This relation allows us to obtain during the  inflationary scenario
 a constraint on the
values of the curvaton field $\sigma_*$ i.e., the value of the curvaton field
 when the cosmological scales exit the horizon. Hence,
from Eq.(\ref{key_2}), at the moment when $H\simeq m$ we get the
restriction

\begin{equation}
\frac{m^2\sigma_*^2}{2\rho_{\phi_{kin}}^{(m)}} \ll
1,\;\;\;\;\mbox{or equivalently}\;\;\;\;
\sigma_*^2\ll\,12 .\label{pot}
\end{equation}
We note that this upper bound for the curvaton field $\sigma_*$
 is similar to that obtained in the standard scalar field
\cite{ureña}.

Also, we find that the ratio between the potential energies at the end of inflation
becomes
\begin{equation}
\frac{U_e}{V_e}\ll 1, \,\,\mbox{or equivalently}\,\,\;\frac{1}{12}\frac{m^2\sigma_*^2}{ H_e^2}\ll
\frac{m^2}{H_e^2 }\label{u},
\end{equation}
 where we considered that  effective potential at the end of inflation is given by
 $V_e=6\, H_e^2$ together with
Eq.(\ref{pot}).  In this way, we find that  the curvaton mass satisfied  the constraint
$m\ll H_e$.
 We  note
that the condition  $m\ll H_e$ is inherent to the
nature of the curvaton field, and becomes a fundamental prerequisite for the
curvaton mechanism\cite{ref1u}.

After  the mass of curvaton field becomes  $m\simeq
H_{kin}$, its energy decays   $\rho_\sigma\propto a^{-3}$ ( non-relativistic matter), and then
we can write
\begin{equation}
\rho_\sigma =
\frac{m^2\sigma_*^2}{2}\frac{a_m^3}{a^3}\label{c_cae}.
\end{equation}

 In the following, we will analyze  the decay of the curvaton
field in two possible different scenarios; the curvaton field decay after domination and the curvaton decay
before domination.

\section{curvaton decay after domination}
As we have required  the curvaton field decay could take place in
two different possible scenarios. During  the first scenario, the
curvaton field decay after domination, and then  the curvaton
field comes to dominates the cosmic expansion, in which
$\rho_\sigma>\rho_\phi$, there must be an instant  when the
inflaton  and curvaton energy densities becomes equivalent.
Considering that both densities becomes equivalent and this occurs
when $a=a_{eq}$, then from Eqs. (\ref{Impor}) and (\ref{c_cae}) we
find
\begin{equation}
\left.\frac{\rho_\sigma}{\rho_{\phi_{kin}}}\right|_{a=a_{eq}}=\frac{m^2\sigma_*^2}{2}\frac{a_m^3}{a_{eq}^3}\,\frac{1}{\dot{\phi_k}^2{\cal
F}_{a_{eq}}^2\left[\frac{A}{2}+\frac{3B}{4}\dot{\phi_k}^2
  {\cal F}_{a_{eq}}^2\right]}
=1,\label{equili}
\end{equation}
where the function ${\cal F}_{a_{eq}}$ corresponds to ${\cal
F}_{a}\mid _{a=a_{eq}}={\cal F}_{a_{eq}}$.

From Eqs.(\ref{c_cae}) and (\ref{equili}), we find a relation for
the Hubble factor, $H(a=a_{eq})=H_{eq}$, as a functions of the
curvaton parameters, together  with  the ratio of the scale factor
at different times

\begin{equation}
H_{eq}= \sqrt{\frac{\sigma_*^2}{12}}\,\,
\left[\frac{a_m}{a_{eq}}\right]^{3/2}\;m. \label{heq}
\end{equation}
Here, we note that this equation coincides with the one found in
standard case, see Refs.\cite{ureña,yo}.

Also, we require  that the curvaton field decays before of
nucleosynthesis, which means $H_{nucl}\sim 10^{-40} <
\Gamma_\sigma$, where the decay parameter $\Gamma_\sigma$ is
constrained by nucleosynthesis. However, we also postulate that
the curvaton decay occurs after $\rho_\sigma > \rho_\phi$,
together with the condition $\Gamma_\sigma < H_{eq}$. In this
form, considering Eq.(\ref{heq}) we can written
\begin{equation}
10^{-40}<\Gamma_{\sigma}<\sqrt{\frac{1}{12}}
\,\,\left[\frac{a_m}{a_{eq}}\right]^{3/2}\,m\,\sigma_*. \label{gamm1}
\end{equation}

Now we will find a constraint of the parameters of our model, by
considering the scalar perturbation associated to the curvaton
field $\sigma$. In this context, the fluctuations of the curvaton
field satisfies an analogous differential equation to the inflaton
fluctuations, and then  we consider that the fluctuations of the
curvaton field takes the amplitude $\delta\sigma_*\simeq
H_*/2\pi$. From the spectrum of the Bardeen parameter
$P_\zeta\simeq H^2_*/(9\pi^2\sigma_*^2)\sim 10^{-9}$ \cite{ref1u},
we get

\begin{equation}
P_\zeta\simeq
\frac{1}{54\pi^2}\frac{V_*}{\sigma_*^2}=\frac{(f_1e^{-\alpha\phi_*}-M_1)^2}{216\pi^2\chi_2
\left[f_2e^{-2\alpha\phi_*}+M_2\right]\,\sigma_*^2},
\label{pafter}
\end{equation}
in which  $\phi_*$ corresponds to the value of the curvaton field
when the cosmological scales exit the horizon. From Ref.[\cite{TMTEMERGENTlast} the
value of $\phi_*$ is given by
$$
e^{-\alpha\phi_*}=\frac{2\alpha
M_1}{f_1(1+bf_1/2f_2)}\,[C_2+2\alpha
N],\,\,\,\;\;\mbox{where}\,\,\,\,\;\;\,C_2=\sqrt{(1+bf_1/2f_2)}.
$$
Here, we have considered that $e^{-\alpha\phi_{e}}=\frac{2\alpha
M_1}{f_1[1+bf_1/f_2]^{1/2}}$, see Ref.\cite{TMTEMERGENTlast}, and
the quantity $N$ corresponds to the number of e-folds.

Considering the constraint given by Eq.(\ref{gamm1}) together with
the condition in which $a_m<a_{eq}$, we get
\begin{equation}
\Gamma_{\sigma}<\sqrt{\frac{1}{12}}
\,\,\left(\frac{m(f_1e^{-\alpha\phi_*}-M_1)^2}{216\pi^2\chi_2P_\zeta\left[f_2e^{-2\alpha\phi_*}+
M_2\right]}\right),\label{G1}
\end{equation}
and we note that this expression gives an upper limit  on the parameter
$\Gamma_\sigma$ when the curvaton decays after domination.

On the other hand, we admit that the reheating occurs before the
bing-bang nucleosynthesis (BBN) temperature $T_{BBN}$, in which
the temperature of reheating $T_{rh}>T_{BBN}$. However, when the
curvaton field decays at the time before the electroweak scale
(ew) where the temperature is $T_{ew}$, then we require that
$T_{rh}\sim\Gamma_{\sigma}^{1/2}>T_{ew}>T_{BBN}$, in which
$T_{ew}\sim 10^{-17}$ and $T_{BBN}\sim 10^{-22}$. In this form,
considering Eq.(\ref{G1}) we can written
\begin{equation}
\frac{m^{1/2}(f_1e^{-\alpha\phi_*}-M_1)}{\chi_2^{1/2}\left[f_2e^{-2\alpha\phi_*}+M_2\right]^{1/2}}>
\left(12\right)^{1/4}(216\,P_\zeta)^{1/2}\,\pi T_{ew}\sim
10^{-20},\label{C1}
\end{equation}
and using Eq.(\ref{m}), the ratio $\mu^{2/3}/\chi_2$ results
\begin{equation}
\frac{\mu^{2/3}}{\chi_2}>10^{-7}.
\end{equation}
Here, following Ref.\cite{TMTEMERGENTlast}  we have considered the
values
 $M_1=4\times10^{-60}$ (in units of $M_{Pl}^4$), $M_2=4$ (in units of $M_{Pl}^4$), $b=-0.52$,
$\alpha=1$, $f_1=2\times10^{-8}$, $f_2=10^{-8}$ and the number of
e-folds $N=60$. In  particular, considering  the range for
$\chi_2$ found in Ref.\cite{TMTEMERGENTlast} in which
$58\times10^{-6}<\chi_2<74\times 10^{-3}$, we obtain  that the
lower bound of the
 parameter $\mu$ becomes $\mu>6\times10^{-13}$. Here, we note that
 from Eqs.(30) and (66), the effective potential $U_+=\frac{M_1^2/M_2}{4\chi_2}\sim
 \frac{10^{-120}}{4\chi_2}> \frac{10^{-198}}{\mu^2}$, and
 considering that the present day value of the density parameter of dark energy  is given by
 $\Omega_{DE}^0=\frac{U_+}{H_0^2}\simeq0.7$ where $H_0$ denotes of the present Hubble parameter $H_0\simeq 10^{-61}$ (in units of
 $M_{Pl}$), we get $\mu>10^{-35}$. In this form, we find a lower bound even smaller for the
 parameter $\mu$, from present dark energy density   in relation to the lower bound for $\mu$, obtained from  the range
 of
$\chi_2$  in Ref.\cite{TMTEMERGENTlast}.
\\

\section{curvaton decay before domination}
In the second scenario the curvaton field decays before it
dominates the cosmological expansion. In this from, we require
that  the curvaton $\sigma$ decays before that its energy density
$\rho_\sigma$ becomes greater than the energy density of the
scalar field $\rho_\phi$. Also, during this scenario the mass of
the curvaton $m$  is similar to the Hubble parameter, i.e., $m\sim
H$, and if the curvaton decays at a moment whenever the decay
parameter $\Gamma_\sigma=H(a_d)=H_d$, in which $d$ label
corresponds to quantities at the moment when the curvaton field
decays. From this condition and considering Eq.(\ref{H}), we get
\begin{equation}
\Gamma_\sigma=H_d=\sqrt{\frac{1}{6}}\;\dot{\phi_k}{\cal
F}_{a_d}\left[
 \frac{A}{2}+\frac{3B}{4}\dot{\phi_k}^2{\cal
 F}_{a_d}^2\right]^{1/2}.
\end{equation}

Before that the curvaton field dominates the cosmological
expansion, i.e., $\rho_\sigma<\rho_\phi$,  the  decay parameter
$\Gamma_\sigma$ satisfies $\Gamma_\sigma>H_{eq}$, and on the other
hand, the curvaton mass $m$ becomes important during this
scenario, in which $m>\Gamma_\sigma$. In this way, considering
Eq.(\ref{heq}) we find
\begin{equation}
\sqrt{\frac{\sigma_*^2}{12}}\,\,
\left[\frac{a_m}{a_{eq}}\right]^{3/2}<\frac{\Gamma_\sigma}{m}<1,\label{c2}
\end{equation}
and these inequalities for the ratio $\Gamma_\sigma/m$, are
similar  to that reported in Ref.\cite{ureña}.

On the other hand, the spectrum of the Bardeen parameter during
this scenario, is given by \cite{Wa}
\begin{equation}
P_\zeta\simeq\frac{r^2_d}{16\pi^2}\,\frac{H^2_*}{\sigma_*^2},\label{sP}
\end{equation}
where the parameter $r_d$ corresponds to the ratio between the
curvaton and inflaton energy densities, evaluated at the moment in
which the curvaton decay takes place i.e., $a=a_d$. From
Eq.(\ref{c_cae}) and considering that $H_d=\Gamma_\sigma$, the
parameter $r_d$ is given by
\begin{equation}
r_d=\left[\frac{\rho_\sigma}{\rho_\phi}\right]_{a=a_d}=\frac{m^2\,\sigma^2_*}{12}\,\left(\frac{a_k}{a_d}\right)^3\,\frac{1}{\Gamma_\sigma^2}.\label{rd}
\end{equation}
Here we remark that the parameter $r_d$ is associated to two other
quantities; the parameter the  non-Gaussianity $f_{NL}$ in which
$f_{NL}\sim r_d^{-1}$, see Ref.\cite{Ko}, and the ratio the
isocurvature and adiabatic amplitudes\cite{St}.

One may ask, is there any smoking gun for this scenario?
The strongest constraints on the non linear parameter $f_{NL}\gg
1$ will get from the measurements of the CMB sky in future
detection, and in this form the non linear parameter would be a
smoking gun for the curvaton-two measures theory.  In particular
if the non Gaussianity is of the local type, the non linear
parameter $f_{NL}^{local}$  from the curvaton field considering
the Planck data 2015 is given by $f_{NL}^{local}=2.5\pm5.7$ at
68$\%$ CL \cite{NG0}, in the case in which there is no import
decay of the inflaton field into curvaton particles. Otherwise, if
the inflaton field into curvaton particles the non linear
parameter $f_{NL}^{local}$ was obtained in Ref.\cite{Sa}, and from
Planck results $f_{NL}^{local}\sim O (1)$. However, due to the
measurement errors no conclusive affirmation can be made, since
the data are preliminaries\cite{NG0}, and therefore  the
observational ranges on the magnitude of the non linear parameter
will progress considerably in the  near future.
\\
\\
On the other hand, another conceivable signature in our model is
the isocurvature perturbation from the baryon or neutrinos density
that is associated  to the curvature perturbation,  from the
curvaton decay. This  isocurvature perturbation was studied in
Refs.[32]\cite{L}, and this mechanism of correlation should be
measurement,  and could become an observable magnitude from CMB
observations\cite{L1}.

Now combining Eqs.(\ref{sP}) and (\ref{rd}) we obtain that the
curvaton field $\sigma_*^2$ can be written as
\begin{equation}
\sigma_*^2=\frac{2304\pi^2\,P_\zeta}{m^4\,H_*^2}\,
\left[\frac{a_d}{a_k}\right]^6\,\Gamma_\sigma^4,\label{s2}
\end{equation}
and by using Eqs.(\ref{c2}) and (\ref{s2}), we find an upper bound
for the parameter $\Gamma_\sigma$ given by
\begin{equation}
\Gamma_\sigma^3<\frac{m^3\,\,H_*^2}{1152\pi^2\,P_\zeta}\left[\frac{a_k}{a_d}\right]^6.\label{G2}
\end{equation}

Now considering that the reheating temperature $T_{rh}$ satisfies
the constraint $T_{rh}>T_{BBN}\sim 10^{-22}$, with
$\Gamma_\sigma>T^{2}_{BBN}$ we find
\begin{equation}
m\,H^{2/3}_*\,\left[\frac{a_k}{a_d}\right]^3>(1152\,\pi^2)^{1/3}\,P_\zeta^{1/3}\,T^2_{BBN}\sim
10^{-44}.
\end{equation}
Considering that $a_d>a_k$, and that the curvature perturbations
generated during inflation  are due to quantum fluctuations of the
curvaton field, where the energy scale is approximately
$V_*^{1/4}\approx10^{15-16}$ GeV (upper bound)\cite{Di}, we get

\begin{equation}
\frac{\mu}{\chi_2^{1/2}}>(6912\,\pi^2)^{1/3}\,\,\frac{\,P_\zeta^{1/3}\,T^2_{BBN}}{V_*^{1/3}}\,\sqrt{\frac{2\,f_2}{f_1}}\sim
10^{-39}.
\end{equation}
Here we have used Eq.(\ref{m}), and the values
$f_1=2\times10^{-8}$  and $f_2=10^{-8}$  from
Ref.\cite{TMTEMERGENTlast}. This expression gives a lower bound on
the parameters $\mu/\chi_2^{1/2}$, during the curvaton field decay before
domination. In  particular, using the range for $\chi_2$ from Ref.\cite{TMTEMERGENTlast}  
in which $58\times10^{-6}<\chi_2<74\times 10^{-3}$, we find  that the lower 
bound of the 
 parameter $\mu$ is given by $\mu>2\times10^{-40}$.

\section{Conclusions \label{conclu}}

In this paper we have studied in detail the curvaton reheating
mechanism in a scale invariant two measures theory.  In this framework 
 the   responsible for the reheating of the universe as well as  the spectrum of
curvature perturbations is the curvaton field $\sigma$.

We have considered that our model involves two scalar matter fields, a dilaton, that
transforms under scale transformations and it drives the expansion of the universe, and 
another scalar, which does not
transform under scale transformations and it play the role
of a curvaton field, with an effective mass during the reheating of the universe.

Considering  the curvaton reheating mechanism we have examined  two possible scenarios. 
Firstly, the curvaton field dominates the universe after it decays, and  in the second scenario
 the curvaton field decays before
domination.  In these scenarios, 
 we have  obtained   constraints for the values of the decay parameter $\Gamma_\sigma$ 
 which are represented
by Eqs.(\ref{G1}) and (\ref{G2}), respectively.

From the stage in which the curvaton field decays after its dominates, we have 
obtained a lower limit for the ratio $\mu^{2/3}/\chi_2>10^{-7}$. 
Here, we have considered Eq. (\ref{C1}),  together with the values found in Ref.\cite{TMTEMERGENTlast} , in which 
 $M_1=4\times10^{-60}$, $M_2=4$, $b=-0.52$,
$\alpha=1$, $f_1=2\times10^{-8}$, $f_2=10^{-8}$ and the number of e-folds
$N=60$. In  particular, considering  the range for $\chi_2$ found in Ref.\cite{TMTEMERGENTlast}  
in which $58\times10^{-6}<\chi_2<74\times 10^{-3}$, we have obtained  that the lower 
bound of the 
 parameter $\mu$ becomes $\mu>6\times10^{-13}$.

In the second scenario, we could estimate the lower bound for the 
ratio $\mu/\chi_2^{1/2}>10^{-39}$. Here we have considered Eq.(\ref{m}), and the values
$f_1=2\times10^{-8}$  and $f_2=10^{-8}$  from
Ref.\cite{TMTEMERGENTlast}. Also, in  particular, using the range for $\chi_2$ from Ref.\cite{TMTEMERGENTlast}  
 we have found  that the lower 
bound of the 
 parameter $\mu$ is given by $\mu>2\times10^{-40}$. 
 
One may ask, is there any smoking gun for this scenario?
The strongest constraints on the non linear parameter $f_{NL}\gg
1$ will get from the measurements of the CMB sky in future
detection, and in this form the non linear parameter would be a
smoking gun for the curvaton-two measures theory.  In particular
if the non Gaussianity is of the local type, the non linear
parameter $f_{NL}^{local}$  from the curvaton field considering
the Planck data 2015 is given by $f_{NL}^{local}=2.5\pm5.7$ at
68$\%$ CL \cite{NG0}, in the case in which there is no import
decay of the inflaton field into curvaton particles. Otherwise, if
the inflaton field into curvaton particles the non linear
parameter $f_{NL}^{local}$ was obtained in Ref.\cite{Sa}, and from
Planck results $f_{NL}^{local}\sim O (1)$. However, due to the
measurement errors no conclusive affirmation can be made, since
the data are preliminaries\cite{NG0}, and therefore  the
observational ranges on the magnitude of the non linear parameter
will progress considerably in the  near future.

In both scenarios the values for the couplings considered allow an emergent non singular scenario followed by inflation  as studied in Ref.\cite{TMTEMERGENTlast}, since the solutions for the emergent phase found in Ref.\cite{TMTEMERGENTlast}
hold as solutions of the model enriched by the curvaton that we have added now in the case $\epsilon = 0$ and neglecting the curvaton (i.e. considering $\sigma = 0$ or negligible in the early emergent phase).

Finally we discuss
generalizations of the model, through the effect of higher
curvature terms,  where inflaton and curvaton can have coupled
oscillations. Indeed, as we have mentioned, for the case of $\epsilon =0$, there are no mixed curvaton -inflaton kinetic terms, but for any non vanishing value of $\epsilon$, those terms will appear, and they will induce inflaton - curvaton oscillations, a subject of interest for future studies.

\begin{acknowledgments}
EG wants to thank the Pontificia Universidad Cat\'olica de
Valpara\'iso for hospitality. R H was supported by Comisi\'on
Nacional de Ciencias y Tecnolog\'ia of Chile through FONDECYT
Grant N$^{0}$ 1130628 and DI-PUCV N$^{0}$ 123.724.

\end{acknowledgments}



\begin{thebibliography}{99}
\bib{early-univ}
E.W. Kolb and M.S. Turner, {\em ``The Early Universe''}, Addison Wesley (1990); \\
A. Linde, {\em ``Particle Physics and Inflationary Cosmology''}, Harwood, Chur,
Switzerland (1990); \\
A. Guth, {\em ``The Inflationary Universe''}, Vintage, Random House (1998); \\
A.R. Liddle and D.H. Lyth, {\em ``Cosmological Inflation and Large-Scale Structure''},
Cambridge Univ. Press (2000); \\
S. Dodelson, {\em ``Modern Cosmology''}, Acad. Press (2003);\\
S. Weinberg, {\em ``Cosmology''}, Oxford Univ. Press (2008).
\bib{primordial}
V. Mukhanov, {\em ``Physical Foundations of Cosmology''}, Cambride Univ. Press (2005); \\
A.R. Liddle and D.H. Lyth, {\em ``The Primordial Density Perturbations -- Cosmology,
Inflation and Origin of Structure''}, Cambridge Univ. Press (2009).
\bib{accel-exp}
M.S. Turner, in \textsl{Third Stromle Symposium ``The Galactic Halo''},
ASP Conference Series Vol.{\bf 666}, B.K. Gibson, T.S. Axelrod and M.E. Putman
(eds.), 1999; \\
N. Bahcall, J.P. Ostriker, S.J. Perlmutter and P.J. Steinhardt, \textsl{Science}
{\bf 284} (1999) 1481;\\
for a review, see P.J.E. Peebles and B. Ratra, {\sl Rev. Mod. Phys.} {\bf 75}
(2003) 559.
\bib{accel-exp-2}
A. Riess, {\em et al.}, \textsl{Astronomical Journal} {\bf 116} (1998) 1009-1038; \\
S. Perlmutter {\em et al.}, \textsl{Astrophysical Journal} {\bf 517} (1999) 565-586.
\bib{peebles-vilenkin}
P.J.E. Peebles and A.Vilenkin, \PRD{59}{1999}{063505}.
\bib{starobinsky-2}
S. Nojiri and S. Odintsov, \PRD{68}{2003}{123512} ~(\textsl{arxiv:hep-th/0307288}); \\
G. Cognola, E. Elizalde, S. Nojiri, S.D. Odintsov, L. Sebastianiand S. Zerbini,
\PRD{77}{2008}{046009} ~(\textsl{0712.4017} [hep-th]), and references therein;\\
S.A. Appleby, R.A. Battye and A.A. Starobinsky,
\textsl{JCAP} {\bf 1006} (2010) 005 ~(\textsl{arxiv:0909.1737} [astro-ph]).
\bibitem{k-essence}
T. Chiba, T.Okabe and M. Yamaguchi, \PRD{62}{2000}{023511}
~(\textsl{arxiv:astro-ph/9912463});\\
C. Armendariz-Picon, V. Mukhanov and P. Steinhardt, \PRL{85}{2000}{4438}
~(\textsl{arxiv:astro-ph/0004134}); \\
C. Armendariz-Picon, V. Mukhanov and P. Steinhardt,
\PRD{63}{2001}{103510} ~(\textsl{arxiv:astro-ph/0006373}); \\
T. Chiba, \PRD{66}{2002}{063514} ~(\textsl{arxiv:astro-ph/0206298}).
\bib{saitou-nojiri}
R. Saitou and S. Nojiri, \textsl{Eur. Phys. J.} {\bf C71} (2011) 1712
~(\textsl{arxiv:1104.0558} [hep-th]).
\bib{wetterich}
C. Wetterich, \PRD{89}{2014}{024005} ~(\textsl{arxiv:1308.1019} [astro-ph]).
\bib{murzakulov-etal}
Md. Wali Hossain, R. Myrzakulov, M. Sami and E.N. Saridakis, \PRD{90}{2014}{023512}
~(\textsl{arxiv:1402.6661} [gr-qc]).
\bibitem{TMTEMERGENTlast} E. Guendelman, R. Herrera, P. Labrana, E. Nissimov and S. Pacheva,
\textsl{Gen. Rel. Grav.} \textbf{47} (2015) 2, 10. (\textsl{arxiv:1408.5344} [gr-qc]).
\bib{TMT-recent-1-a}
S. del Campo. E. Guendelman, R. Herrera and P. Labrana, \textsl{JCAP} {\bf 1006}
(2010) 026 ~(\textsl{arxiv:1006.5734} [astro-ph.CO]).
\bib{TMT-recent-1-c}
E.I. Guendelman and P. Labrana, \IJMPD{22}{2013}{1330018}
~(\textsl{arxiv:1303.7267} [astro-ph.CO]).

\bib{TMT-orig-1}
E.I. Guendelman, \MPLA{14}{1999}{1043-1052} ~(\textsl{arxiv:gr-qc/9901017});\\
E.I. Guendelman, in {\em ``Energy Densities in the Universe''}, Proc.
Rencontres de Moriond, Les Arcs (2000) ~(\textsl{arxiv:gr-qc/0004011}).
\bib{TMT-orig-2}
E.I. Guendelman and A. Kaganovich, \PRD{60}{1999}{065004}
~(\textsl{arxiv:gr-qc/9905029}).
\bib{TMT-orig-3}
E.I. Guendelman and O. Katz, \CQG{20}{2003}{1715-1728}
~(\textsl{arxiv:gr-qc/0211095}).
\bib{TMT-recent-2}
E.I. Guendelman, D. Singleton and N. Yongram, \textsl{JCAP} {\bf 1211} (2012) 044
~(\textsl{arxiv:1205.1056} [gr-qc]);\\
E.I. Guendelman, H. Nishino and S. Rajpoot, \PLB{732}{2014}{156}
~(\textsl{arxiv:1403.4199} [hep-th]).
\bib{nishino-rajpoot}
H. Nishino and S. Rajpoot, \PLB{736}{2014}{350-355}
~(\textsl{arxiv:1411.3805} [hep-th]).
\bib{mstring}
E. Guendelman,  \CQG{17}{2000}{3873-3880}, (\textsl{arxiv:hep-th/0005041}),
E. Guendelman, \PRD{63}{2001}{046006}
~(\textsl{arxiv:hep-th/00051232).
E. Guendelman, A. Kaganovich, E. Nissimov and S. Pacheva, \PRD{66}{2002}{046003}}
~(\textsl{arxiv:hep-th/0203024}).
\bib{susy-break}
E. Guendelman, E.Nissimov, S. Pacheva and M. Vasihoun, \textsl{Bulg. J. Phys.}
{\bf 40} (2013) 121-126 ~(\textsl{arxiv:1310.2772} [hep-th]); \\
E. Guendelman, E.Nissimov, S. Pacheva and M. Vasihoun, \textsl{Bulg. J. Phys.}
{\bf 41} (2014) 123-129 ~(\textsl{arxiv:1404.4733} [hep-th]).
\bib{Struckmeier}
J. Struckmeier, \PRD{91}{2015}{085030}
~(\textsl{arxiv:1411.1558}[gr-qc]).
\bib{starobinsky}
A. Starobinsky, \PLB{91}{1980}{99-102}.
\bib{zee-induced-grav}
A. Zee, \PRL{42}{1979}{417}.

\bib{TMT-recent-1-b}
S. del Campo. E. Guendelman, A. Kaganovich, R. Herrera and P. Labrana,
\PLB{699}{2011}{211} ~(\textsl{arxiv:1105.0651} [astro-ph.CO]).

\bib{emergent-univ}
G.F.R. Ellis and R. Maartens, \CQG{21}{2004}{223} ~(\textsl{gr-qc/0211082}); \\
G.F.R. Ellis, J. Murugan and C.G. Tsagas,\CQG{21}{2004}{233}
~(\textsl{ arxiv:gr-qc/0307112});\\
D.J. Mulryne, R. Tavakol, J.E. Lidsey and G.F.R. Ellis,
\PRD{71}{2005}{123512} ~(\textsl{arxiv:astro-ph/0502589}); \\
A. Banerjee, T. Bandyopadhyay and S. Chaakraborty, {\sl Grav. Cosmol.} {\bf 13}
(2007) 290-292 ~(\textsl{arxiv:0705.3933} [gr-qc]); \\
J.E. Lidsey and D.J. Mulryne \PRD{73}{2006}{083508}
~(\textsl{arxiv:hep-th/0601203}); \\
S. Mukherjee, B.C.Paul, S.D. Maharaj and A. Beesham, \textsl{arxiv:qr-qc/0505103}; \\
S. Mukherjee, B.C.Paul,  N.K. Dadhich, S.D. Maharaj and A. Beesham,
\CQG{23}{2006}{6927} ~(\textsl{arxiv:gr-qc/0605134}).
\bibitem{see saw} 
M. Gell Mann, P. Ramond and R.Slansky, in Supergravity, edited by D. Friedman (North Holland,
Amsterdam, 1979), p. 315, T. Yanagida in Proceedings of the workshop pn 'Unified Theories and Baryon Number in the Universe', edited by O. Sawada and A Sugamoto (KEK, Tsukuba, Japan, 1979), 
R. Mohapatra and G. Senjanovich, Phys. Rev. Lett. 44, 912 (1980) and Phys. Rev.23, 165 (1981),
A.Davidson and K. Wali, Phys. Rev. Lett. 59, 393 (1987), 
S. Rajpoot, Phys.Lett. B191 (1987) 122, 
S. Rajpoot, Phys.Rev. D36 (1987) 1479-1483,  
S. Rajpoot,  Mod.Phys.Lett. A2 (1987) 5, 307-315,S. Rajpoot, Mod.Phys.Lett. A2 (1987) 75, 541.




\bibitem{ureña}A. R. Liddle and L. A. Ure\~na-L\'opez,
Phys. Rev. D{\bf 68}, 043517 (2003).



\bibitem{ref1u}D. H. Lyth and D. Wands, Phys. Lett. B{\bf 524}, 5
(2002); S. Mollerach, Phys. Rev. D {\bf 42}, 313 (1990).

\bibitem{Feng-li}
Bo Fend and Mingzhe Li, \PLB{564}{2003}{169-174}(\textsl{arxiv:hep-ph/0212213}).

\bibitem{Campuzano:2005qw} 
  C.~Campuzano, S.~del Campo,  and R.~Herrera,
  Phys.\ Rev.\ D {\bf 72}, 083515 (2005)
  [Phys.\ Rev.\ D {\bf 72}, 109902 (2005)];
  C.~Campuzano, S.~del Campo and R.~Herrera,
  Phys.\ Lett.\ B {\bf 633}, 149 (2006).
  
  \bibitem{yo}C.~Campuzano, S.~del Campo and R.~Herrera,
  JCAP {\bf 0606}, 017 (2006); H.~S.~Zhang and Z.~H.~Zhu,
  Phys.\ Lett.\ B {\bf 641}, 405 (2006); 
  S.~del Campo and R.~Herrera,
  Phys.\ Rev.\ D {\bf 76}, 103503 (2007);
  S.~del Campo, R.~Herrera, J.~Saavedra, C.~Campuzano and E.~Rojas,
  Phys.\ Rev.\ D {\bf 80}, 123531 (2009).

\bibitem{Now} C.~He, D.~Grin and W.~Hu,
  arXiv:1505.00639 [astro-ph.CO]; R.~J.~Hardwick and C.~T.~Byrnes,
  arXiv:1502.06951 [astro-ph.CO]; K.~Feng and T.~Qiu,
  Phys.\ Rev.\ D {\bf 90}, no. 12, 123508 (2014); S.~Clesse, B.~Garbrecht and Y.~Zhu,
  JCAP {\bf 1410}, no. 10, 046 (2014); K.~Feng, T.~Qiu and Y.~S.~Piao,
  Phys.\ Lett.\ B {\bf 729}, 99 (2014).
  

\bibitem{Wa}D. Lyth, C. Ungarelli and D. Wands, Phys. Rev. D {\bf
67}, 023503 (2003).

\bibitem{Ko} E. Komatsu and D. Spergel, Phys. Rev. D
{\bf63}, 063002 (2001).

\bibitem{NG0} P.~A.~R.~Ade {\it et al.} [Planck Collaboration],
  arXiv:1502.01592 [astro-ph.CO].
  
\bibitem{Sa}
 M. Sasaki, J. Valiviita, and D. Wands,  Phys.Rev.,
D{\bf74}, 103003 (2006).

\bibitem{L}
  D.~H.~Lyth and Y.~Rodriguez,
  Phys.\ Rev.\ Lett.\  {\bf 95}, 121302 (2005);J.~Lesgourgues and S.~Pastor,
  Phys.\ Rept.\  {\bf 429}, 307 (2006).

\bibitem{L1}E.~Di Valentino, M.~Lattanzi, G.~Mangano, A.~Melchiorri and P.~Serpico,
  Phys.\ Rev.\ D {\bf 85}, 043511 (2012);
 K.~Harigaya, T.~Hayakawa, M.~Kawasaki and S.~Yokoyama,
  JCAP {\bf 1410}, no. 10, 068 (2014).


\bibitem{St} R. Stompor, A. Banday and K. Gorski, Astrophys. J. {\bf
463}, 8 (1996).


\bibitem{Di} K. Dimopoulos and D. Lyth, Phys. Rev. D
{\bf69}, 123509 (2004).






























\end{thebibliography}
\end{document}